\documentclass[conference]{IEEEtran}
\IEEEoverridecommandlockouts

\usepackage{cite}
\usepackage{amsmath,amssymb,amsfonts}
\usepackage{graphicx}
\usepackage{textcomp}
\def\BibTeX{{\rm B\kern-.05em{\sc i\kern-.025em b}\kern-.08em
    T\kern-.1667em\lower.7ex\hbox{E}\kern-.125emX}}

\usepackage{centernot}
\usepackage{subfig}
\usepackage[inline]{enumitem}
\usepackage[linesnumbered,commentsnumbered,ruled,vlined]{algorithm2e}
\usepackage{framed}
\usepackage{semantic}
\usepackage[final]{listings}
\usepackage{caption}
\usepackage{hyperref}

\usepackage{bbold}
\usepackage{tablefootnote}
\usepackage{multirow}
\usepackage{booktabs}
\usepackage{balance}
\newtheorem{example}{\textbf{Example}}
\newtheorem{definition}{\textbf{Definition}}
\newtheorem{theorem}{\textbf{Theorem}} 
\newtheorem{proof}{\textbf{Proof}}
\newtheorem{lemma}{\textbf{Lemma}}

\usepackage{xargs}
\usepackage[colorinlistoftodos,prependcaption]{todonotes}
\newcommandx{\unsure}[2][1=]{\todo[inline,linecolor=red,backgroundcolor=red!25,bordercolor=red,#1]{#2}}
\newcommandx{\change}[2][1=]{\todo[inline,linecolor=blue,backgroundcolor=blue!25,bordercolor=blue,#1]{#2}}
\newcommandx{\info}[2][1=]{\todo[inline,linecolor=OliveGreen,backgroundcolor=OliveGreen!25,bordercolor=OliveGreen,#1]{#2}}
\newcommandx{\improvement}[2][1=]{\todo[inline,linecolor=orange,backgroundcolor=orange!25,bordercolor=orange,#1]{#2}}
\newcommandx{\thiswillnotshow}[2][1=]{\todo[disable,#1]{#2}}

\usepackage{soul}

\usepackage{pgfplots}
\pgfplotsset{compat=1.10}
\usepackage{pifont}

\usepackage[authormarkup=none]{changes}

\definechangesauthor[color=purple]{xiaoning}

\definecolor{codegreen}{rgb}{0.0, 0.0, 1.0}

\usepackage{listings}
\lstdefinelanguage{sql}
{alsoletter={-, =}, morekeywords={SELECT, FROM, WHERE, GROUP, BY, UNION, ALL,
	SUM, TRUE, FALSE, NULL}, sensitive=false, morecomment=[l]{;}}
\usepackage{tcolorbox}
\newcommand{\mybox}[1]{
	\begin{tcolorbox}[boxsep=-0.5pt, standard jigsaw, boxrule=0.6pt,
		opacityback=0, sharp corners] #1
	\end{tcolorbox}
}

\usepackage{tikz}
\usetikzlibrary{arrows, automata}

\usepackage[utf8]{inputenc}
\usepackage{cleveref}
\crefname{section}{§}{§§}
\Crefname{section}{§}{§§}

\begin{document}

\title{Duplicate-sensitivity Guided Transformation Synthesis for DBMS Correctness Bug Detection}

\author{
\IEEEauthorblockN{Yushan Zhang}
\IEEEauthorblockA{\textit{Department of Computer Science and Engineering} \\
\textit{HKUST}\\
Hong Kong, China \\
yzhanghw@connect.ust.hk} \\
\IEEEauthorblockN{Rongxin Wu}
\IEEEauthorblockA{\textit{Department of Cyber Space Security} \\
\textit{School of School of Informatics} \\
\textit{Xiamen University}\\ 
Xiamen, China \\
wurongxin@xmu.edu.cn}
\and
\IEEEauthorblockN{Peisen Yao}
\IEEEauthorblockA{\textit{Department of Computer Science and Engineering} \\
\textit{HKUST}\\
Hong Kong, China \\
pyao@cse.ust.hk} \\
\IEEEauthorblockN{Charles Zhang}
\IEEEauthorblockA{\textit{Department of Computer Science and Engineering} \\
\textit{HKUST}\\
Hong Kong, China \\
charlesz@cse.ust.hk}
}

\maketitle

\begin{abstract}
  Database Management System (DBMS) plays a core role in modern software from
  mobile apps to online banking. 
  It is critical that the DBMS provides correct data to all applications. 
  When the DBMS returns incorrect data, a correctness bug is triggered.
  Current production-level DBMSs still suffer from insufficient testing due
  to the limited hand-written test cases. Recently several works proposed to
  automatically generate many test cases with query transformation, a process of
  generating an equivalent query pair and testing a DBMS by checking whether the
  system returns the same result set for both queries. However, all of them still
  heavily rely on manual work to provide a transformation which largely confines
  their exploration of the valid input query space. 
  
  This paper introduces
  duplicate-sensitivity guided transformation synthesis which automatically
  finds new transformations by first synthesizing many candidates then filtering
  the nonequivalent ones. Our automated synthesis is achieved by mutating a
  query while keeping its duplicate sensitivity, which is a necessary condition
  for query equivalence. After candidate synthesis, we keep the mutant query
  which is equivalent to the given one by using a query equivalent checker.
  Furthermore, we have implemented our idea in a tool Eqsql and used it to test
  the production-level DBMSs. In two months, we detected in total 30 newly
  confirmed and unique bugs in MySQL, TiDB and CynosDB.
  
\end{abstract}


\section{Introduction}
\label{sec:intro}
Database Management Systems (DBMSs) are widely used in modern industries. The
users expect the DBMS to retrieve correct data records and report correct
analysis results on the giving databases. Unexpectedly, researchers could still
find hundreds of queries  where the popular and production-level DBMSs return
wrong result sets~\cite{Rigger2020PQS}, though these systems have been
extensively tested during their
development~\cite{wb-howsqlite,wb-mysqltest}.

To validate the correctness of different DBMS components (e.g. the optimizer),
the developers usually  manually write test cases and provide the expected
result for a query. However, the quality of test cases is largely limited by
developers' testing skills~\cite{11factors} and the hand-written test cases only
cover a very small part of the possible input query space~\cite{08Microsoft},
leading to insufficient testing of the DBMS. In what follows, we refer to a test
case for the purpose of DBMS correctness testing as a test query and an expected result set.

\begin{figure*}[t] 
	\centering
		\includegraphics[scale=0.50]{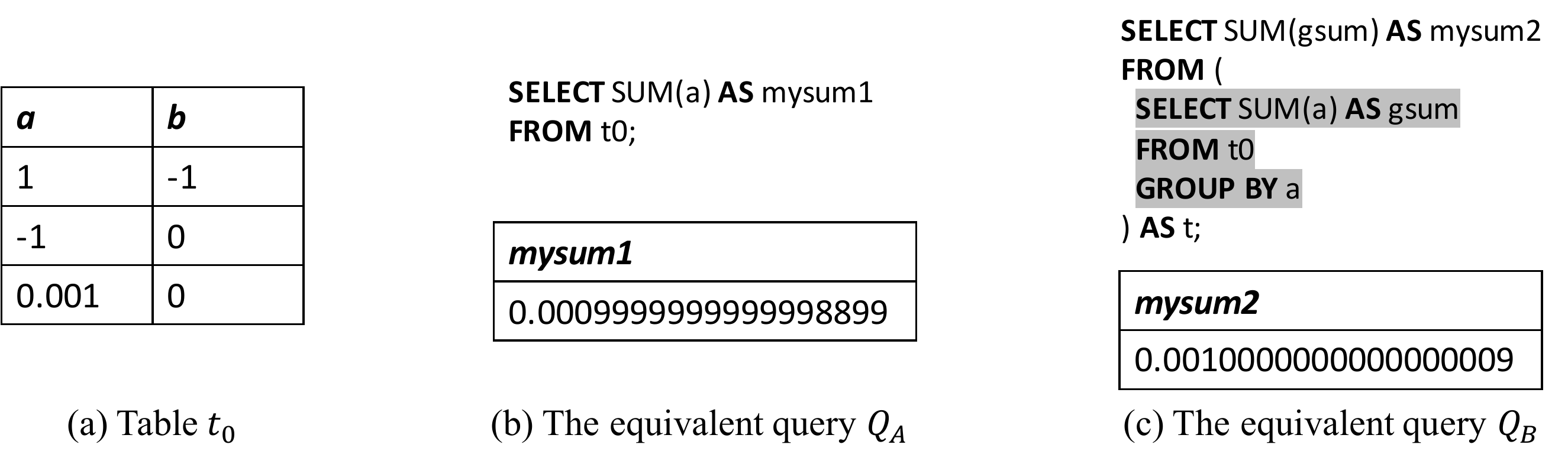}
  	\caption{$Q_A$ and $Q_B$ is an equivalent query pair, which is expected to
  	compute the same results. In MySQL they produce different results, which
  	revealed an inconsistent float representation bug.} \label{fig:casestudy}
\end{figure*}

Recent studies~\cite{08adusa,veanes2010qex,20norec,20mutasql,Rigger2020TLP} have
proposed several approaches to better explore the query space. They all use
automated random query generation to enlarge the explored query space. However, the
key challenge is the test oracle problem, i.e., how to find the expected result
set of a given query. To address the problem, one category of approaches such as
ADUSA~\cite{08adusa} and Qex~\cite{veanes2010qex} convert the query to Satisfiability Modulo Theory (SMT) 
constraints and use a constraint solver to deduce the results. However, they
cannot work on special data types (e.g. varchar) in the DBMS and big tables
because of the performance issue of the solver. Another category of approaches
identify a correctness bug by comparing the result set of a pair of equivalent
queries. However, existing approaches can only use limited rules/transformations
to derive the query pairs. For example, TLP~\cite{Rigger2020TLP} first generates
a query $Q$ and then uses a predefined transformation to on $Q$ to obtain a new
equivalent query $Q^\prime$. Next, it checks whether the DBMS generates the same
result under the input of $Q$ and $Q^\prime$. However, since some correctness
bugs can only be exposed under a certain query pair and the predefined
transformation may fail to generate such query pair, these approaches would lose
the chances to detect such bugs. To be more specific, the transformation rule of
the existing studies \cite{Rigger2020PQS,20norec,Rigger2020TLP}
presumes that the generated query pair should explicitly include a ``where"
predicate (e.g. where a != 0) and the bug was triggered by the inconsistent
interpretations of ``where'' predicates. However, as shown in
\autoref{fig:casestudy}, this bug is exposed by two equivalent queries $Q_A$
(Fig 6b) and $Q_B$ (Fig 6c) in which there are no ``where'' predicates.
Essentially, this bug is caused by an inconsistent representation of the
float-point value $0.001$. $Q_A$ and $Q_B$ are expected to return the same value
on the given table t0, but they have a different value for $0.001$ as shown in
the result column.
\

The key point in identifying such a correctness bug in \autoref{fig:casestudy} is to
compare if a row appears in the result sets of the equivalent query pair. 
In this example, \texttt{mysum1} and \texttt{mysum2} do not have a same row, 
and then this bug is exposed. 
Inspired by this example, 
our key idea is to create such a query that a row appears the same number of times 
in the result sets of both the original and transformed queries.
To achieve this goal, it is required to describe the property whether duplicated
rows should present in a query's result set, 
which we denote as \textit{query duplicate sensitivity}.
This property is generally
available on all SQL queries, which does not require the query to have a
specific component such as the ``where'' predicate. Although we can create
query mutants by preserving this property, it can generate queries which are not
equivalent to the given one because it is a necessary condition. To keep
the valid query mutants only, we leverage the existing query equivalence
checker~\cite{zhou2019automated} to
do the job.

Following our insight, we designed an automated correctness bug detection tool with duplicate-sensitivity
guided query transformations. Eqsql could synthesize many candidate
mutant transformations for a query and it uses the query equivalent solver EQUITAS
~\cite{zhou2019automated} to identify the valid mutants. To evaluate the effectiveness of our approach, we
have implemented it in a tool called Eqsql. We ran Eqsql on several real-world
production-level DBMSs, namely MySQL \cite{wb-mysql}, PingCAP TiDB
\cite{wb-tidb} and Tecent CynosDB \cite{wb-cdb}. During a two month evaluation,
our tool has found 30 confirmed and unique bugs. It detected 14 bugs in
MySQL, 13 bugs in TiDB and 3 bugs in Tencent CynosDB. While we evaluated our
tool on MySQL-compatible database systems, we believe our approach is also
applicable to the other DBMSs.

\definecolor{anti-flashwhite}{rgb}{0.95, 0.95, 0.96}

\definecolor{whitesmoke}{rgb}{0.96, 0.96, 0.96}

\smallskip
In summary, we made the following contributions:
\begin{itemize}
	\item We present a workflow for synthesizing equivalent-preserving query
	transformations, based on duplicate sensitivity of SQL clauses.

	\item We implemented our approach in Eqsql, which is a fully automated
	testing tool for detecting correctness bugs in MySQL-compatible database
	systems. 
	
	\item We detected 30 new and unique bugs in the popular production-level
	DBMSs, as a practical evaluation of Eqsql.
\end{itemize}

%
\section{Illustrative Examples}

In this section, we use two examples to demonstrate how we obtain more
transformations on a given query. Before we look into the examples, we first
need to define some jargon to help use explain concepts in the SQL language. 
\begin{itemize}
	\item \textbf{SQL Operators}. We consider every SQL clause and function as
	a SQL operator. In this section, we introduce the
	\texttt{WHERE}, \texttt{HAVING} and \texttt{DISTINCT} operators.
	\item \textbf{Duplicate Sensitivity}. Every SQL operator must be either
	duplicate sensitive or duplicate insensitive. \texttt{SUM} is duplicate sensitive 
	and \texttt{MAX} is duplicate
	insensitive. Additionally, \texttt{WHERE} and \texttt{HAVING} are duplicate
	sensitive and the \texttt{DISTINCT} operator is duplicate insensitive.
\end{itemize}

\begin{figure}[htbp]
	\centering
	\includegraphics[scale=0.46]{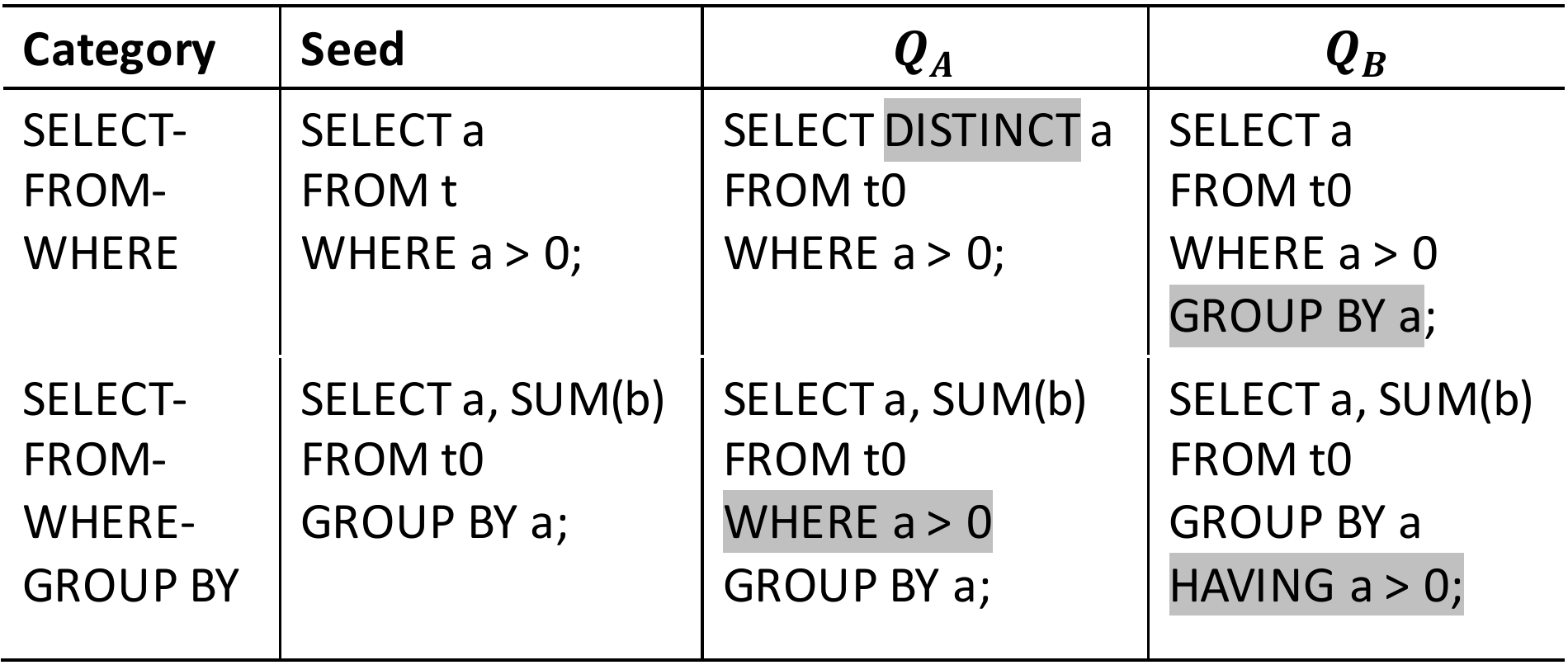}
	\caption{Two SQL categories and seed with generated EQP}
	\label{fig:motiv}
\end{figure}

\smallskip
Due to the limited space, we show how our approach works with two common query
categories: \texttt{SELECT-FROM-WHERE} and \texttt{SELECT-FROM-GROUP BY}. Let us start with the
first category, and then move to the more complex category with aggregations.
\autoref{fig:motiv} shows the categories, the corresponding seed query and the
generated query mutant. The gray color highlights where the transformation takes place.

\section{Category 1: SELECT-FROM-WHERE}
This category is one of the most frequent queries in the DBMS, which asks
the system to filter out data records not satisfying the ``where" predicate.
These queries will go through complex equivalent transformations during the
optimization phase~\cite{98overview}, potentially resulting in a nonequivalent final
result set due to implementation bugs~\cite{09microsoft,08Microftws}. However,
such bugs are hard to be captured due to the large amount of optimization
combinations~\cite{09microsoft}. Previous work NoREC~\cite{20norec} can only
detect such bugs because it only manipulates the ``where" predicate to
create an unoptimized query mutant.

For queries in this category, there are many candidate transformations to apply.
In this example, our approach inserts additional operators to the seed query. By
adding \texttt{DISTINCT} to \texttt{SELECT a} in (\autoref{fig:motiv}), it generated $Q_A$. And by adding groups
with \texttt{GROUP BY}, it gets $Q_B$. Both inserted operators remove duplicates of an
$a$ value and keep only one copy of each different $a$ value. Because adding
duplicates for an $a$ value or remove some duplicates will not change the result
for \texttt{DISTINCT} and \texttt{GROUP BY}, they are \textit{duplicate insensitive}. The
operators de-duplicate the input tuples and output results with one copy for
each row.
We denote this operation as $\delta$, following the convention
in~\cite{Dayal1982}.

\section{Category 2: SELECT-FROM-GROUP BY}
For this category, we consider a query with grouping and an aggregate function.
Different from the first category, queries of this category always produce an
intermediate result~\cite{Guagliardo2017}, which is a set of distinct groups
formed by the \texttt{GROUP BY} columns. A \texttt{GROUP BY} operator splits
the input records into groups based on the given grouping condition. After a
grouping, normally an aggregate function is used to compute some results on a
group, such as a sum of a numeric column. However, we could not extract the
groups explicitly with any SQL operators, which makes it harder to discover such
bugs only with the final result set.

Given an input query with \texttt{SUM} and \texttt{GROUP BY}, we can add a
filter clauses either by inserting a \texttt{WHERE} operator before the
\texttt{GROUP BY} or by inserting a \texttt{HAVING} after it. \texttt{GROUP BY}
is also a de-duplicating operator, because given one copy of each $a$ value is
sufficient to form the groups in our seed query. When both newly-inserted filter
operators \texttt{WHERE} and \texttt{HAVING} apply to $a$, they do not modify
the intermediate groups. Therefore the query mutant $Q_B$ is still equivalent
after the transformation. We denote the filtering operations as $\sigma$.

\smallskip
In this section, we explain the core concepts intuitively with two concrete
queries. Next, we formally describe the scope of SQL syntax and
the problem background in this paper.

\section{Preliminaries}\label{sec:pre}

In this section, we first define the SQL
syntax, followed by a recap of the fundamental
relational data model for the Relational Database Management Systems (RDBMSs). 
We then take special care to describe
aggregation operators due to their different semantic.
Last we describe the key technical challenge addressed in this work.

\subsection{SQL Syntax and Relational Model}
We consider the SQL syntax given in \cite{Guagliardo2017} which supports
aggregate functions and groupings.

\smallskip{\textbf{Notations}}. $n$ is a table name, and $T$ is a set of table
names. $c$ is a column name annotated with the name of the containing table, and
$C$ is a set of column names. A truth value $l$ is an element of the truth
values $L=\{TRUE, FALSE, NULL\}$ in three-valued Logic~\cite{16libkintvl}. $V$
is a set of values which could be assigned to a field. We
use the following symbols: 
\begin{enumerate*}
	\item $T^\prime$ is a sequence of table names.
	\item $C^\prime$ is a sequence of column names.
	\item $t$ is a term, where $t \in L \bigcup V \bigcup C$.
\end{enumerate*}

\autoref{def:syntax} presents the syntax. A query $Q$ could have a ``where''
predicate, denoted as $p$. The predicate can be composed of truth values, an
expression of values, and can be recursively constructed with predicates
connected by binary and unary operators. A query could also use aggregate
functions \texttt{A} in the SELECT part. We also defined \textit{where
predicate} and \textit{operator} to ease our references to specific SQL
structures.

\smallskip
\begin{definition}[Where Predicate]
Given a query, we refer to the part, between \texttt{WHERE} and the next
keyword, as the \textit{where predicate}, or simply as the \textit{predicate}
$P$.
\end{definition}

\smallskip
\begin{definition}[Operator]{\label{def:operator}}
An operator operates on inputs with a type of relation or value (single attribute and single tuple relation), and produces a
result corresponding to the input type. The operator has two categories: 1)
relational operators, which are the operators of a SQL query: \texttt{SELECT,
DISTINCT, WHERE, GROUP BY, UNION, UNION ALL}; 2) value operators, which
manipulate a value, such as aggregate functions and logic operators.
\end{definition}

\begin{figure}[t]
    \centering
    \begin{lstlisting}
        Q := SELECT C' FROM T'
           | SELECT C' FROM T WHERE p
           | SELECT A FROM T'
           | SELECT A FROM T' GROUP BY C
    \end{lstlisting}
    \begin{lstlisting}
        p := TRUE | FALSE | NULL
           | p op p
           | op p
           | v
    \end{lstlisting}
    \caption{SQL syntax discussed in this paper.}
    \label{def:syntax}
\end{figure}

\smallskip{{\textbf{Relational Model}}. Modern RDBMSs use SQL, a language
manipulating table containing duplicates. This motivated us to adopt the extended
relational model in~\cite{Dayal1982}, where Dayal proposed the relational model
over multiset, allowing duplicated values to present in the database.

\smallskip
\begin{definition}[Relation Scheme]
A relation scheme $S$ is a set of $k$-attributes $\{A_1, A_2 \cdots A_k\}$. The
cardinality of the set is the $arity$ of the scheme, denoted as $k$. Each
attribute $A_i$ is associated with its domain: $dom(A)$. A relation scheme could
be written as $S(A_1:dom(A_1),A_2:dom(A_2) \cdots A_k:dom(A_k))$.
\end{definition}

\smallskip
\begin{definition}[Multiset Relation]
A multiset relation $R$ over a relation scheme $S$ is a finite set of pairs
$<t,i>$, where $t$ is a $k$-tuple over $S$ and $i$ is the multiplicity of $t$.
When $i=1$ for every $t$ over $S$, $R$ is a (set-)relation.
\end{definition}

\subsection{Aggregations}
We call the classical aggregate function operators: \texttt{COUNT, SUM, MIN,
MAX, AVG} and the operator \texttt{GROUP BY} as \textit{aggregations (aggs)},
and we refer to a SQL query with any aggs as an \textit{agg query}. Similarly, the
SQL query without aggs is called the \textit{non-agg} query. We consider agg
operators to be different because their results can contain tuples that are not
part of the input relation.

\textbf{Group By}. Hella etc.~\cite{hell99logics} pointed out this operator
produces an intermediate result, which is a set of grouped tuples.
As as long as we use the current SQL syntax, we cannot extract this
intermediate result explicitly with a query.

\textbf{Aggregate Function}. These functions always run on a group, which can
be a result of a \texttt{GROUP BY}. They always produce an one-attribute one-tuple
relation as the result for each and every group.

\subsection{Problem Statement}
At high level, our approach tests a DBMS as follows.
Given a seed query $Q$, we first generate many query mutants by performing different transformations on $Q$. We then run the pair formed by $Q$ with one of the mutant on the tested DBMS. If the DBMS does not produce same result set for them, then a correctness bug is detected.

To transform a SQL query,
we can walk over the production rules of the SQL's context-free grammar, as in conventional grammar-based testing.
However, randomly applying the production rules is inefficient in generating both
syntactically and semantically valid queries~\cite{zhong2020squirrel}, and it can produce many nonequivalent query pairs.
Thus, in this work, the key technical challenge we address is as follows:
\mybox{Given a query $Q_A$, automatically find transformations to
generate candidate mutants $Q_B$ etc., such that 
$Q_A$, $Q_B$ are equivalent.}
}

The cornerstone of our solution is to use the property of 
\textit{duplicate sensitivity} to guide the generation of candidate query mutants.
In the following section, we present the theoretical foundation and the technical detail of our approach.

\section{Approach}\label{sec:approach} 

We start this section with a formal definition for duplicate sensitivity, and
then prove the preservation of duplicate sensitivity as a necessary condition
for query equivalence. Next, 
we present how to  use duplicate sensitivity  to guide the creation of equivalent transformations.
 At last we give the overall algorithm for testing the DBMS.

\subsection{Duplicate Sensitivity for Query Equivalence}

\textbf{Duplicate Sensitivity}. We start by recapping the property of duplicate sensitivity for a single SQL operator such as
\texttt{SUM} and \texttt{MAX}. 
We then extend the notion of duplicate
sensitivity to a whole SQL query. 

\smallskip
\begin{definition}[Duplicate Sensitive Operators]{\label{def:sens}} An operator
is duplicate sensitive, if multiplicity of input changes the result, vice versa.
\end{definition}

\begin{example}
For relational operators, we know that \texttt{SELECT}, \texttt{WHERE} and
\texttt{UNION ALL} are duplicate sensitive and \texttt{DISTINCT}, \texttt{GROUP
BY} and \texttt{UNION} are duplicate insensitive. For value operators we know
that \texttt{MIN}, \texttt{MAX} are insensitive, and \texttt{SUM},
\texttt{COUNT}, \texttt{AVG} are duplicate sensitive.
\end{example}


\smallskip
We have also observed that all operators take a relation as input
(\autoref{def:operator}), which is then processed by the next operator. From
this point of view, a SQL query is a series of operators, which can be either
duplicate sensitive or insensitive.
Thus, we generalize the notion of duplicate sensitivity to a SQL query, defined as below.
\smallskip
\begin{definition}[Query Duplicate Sensitivity]
The duplicate sensitivity of a SQL query is the \textit{ordered composition}
of the duplicate sensitivity of its operators.
\end{definition}

Let $g$ and $h$ be two SQL operators, and $g \cdot h$ be
their composition.
We can obtain the duplicate sensitivity of $g \cdot h$ (denoted as ``$DS(g \cdot
h)$'') as the following:

\begin{equation} \label{eqtime}
	DS(g \cdot h) = \left\{
	\begin{aligned}
	   	\text{sensitive}, &  \ \ g: \text{sensitive} \wedge h: \text{sensitive} \\
		\text{insensitive}, & \ \ others \\
	\end{aligned}
	\right.
\end{equation}
Given any SQL query, we can compute the duplicate sensitivity of it using the above rules.

\smallskip
\begin{example}
Suppose we have a query \texttt{SELECT a FROM t0 WHERE a > 1 GROUP BY a}, we
know the sensitivity for each operator in this query: \texttt{SELECT} is
sensitive, \texttt{WHERE} is sensitive and \texttt{GROUP BY} is insensitive. Formally, each operator is a transformation function which is given in
\autoref{fig:dsl}. The ordered composition of these operators is applying
these functions in order.
One possible composition of the query is $\pi \cdot \delta \cdot \sigma$, and the ordered composition of
its operators' sensitivity can be derived by following \autoref{eqtime}, which is insensitive.

\end{example}

\begin{figure}[t]
		\begin{align}
			\pi : \tau \rightarrow \tau \tag{SELECT} \label{type:pi} \\
			\sigma : \tau \rightarrow \tau \tag{WHERE, HAVING} \label{type:sigma} \\
			\gamma : \tau \rightarrow \tau_v \tag{Aggregate Function} \\
			\delta : \tau_b \rightarrow \tau_s \tag{DISTINCT, GROUP BY}
		\end{align}
	\caption{Describing SQL clause input and output. $\tau_b$ is a multiset
		type, $\tau_s$ is a set type, $\tau_v$ is a value type and 
		$\tau$ is generic it could be either of the three. A set type is duplicate insensitive
		, a multiset type is duplicate sensitive and a value type's sensitivity depends on the
		function.}
	\label{fig:dsl}
\end{figure}

\smallskip{\textbf{Necessary Condition for Query Equivalence}}. We now prove
that two equivalent queries must have the same duplicate sensitivity by
constructing a contrapositive proof using query equivalence.

\begin{lemma}[Query Equivalence]{\label{def:equiv}} Given two queries $Q_1$ and
$Q_2$. They are equivalent, denoted as $Q_1 \equiv Q_2$ if and only if the
following holds:

$\forall D, R_1(D) \subseteq  R_2(D) \bigwedge R_2(D) \subseteq R_1(D)$. 

$R_1(D)$ is the result set of $Q_1$ on relation $D$ and $R_2(D)$ is the result set of
$Q_2$ on relation $D$, respectively.
\end{lemma}

\smallskip
\begin{theorem}{\label{thm:dsth}} Given two queries $Q_1$ and $Q_2$, if $Q_1
\equiv Q_2$, then $DS(Q_1) \equiv DS(Q_2)$. Where $DS(Q_1)$ and $DS(Q_2)$ denote
\textit{duplicate-sensitivity} of $Q_1$ and $Q_2$, respectively.
\end{theorem}

\smallskip
\begin{proof}
We construct a contrapositive proof for this theorem. The theorem when formally
written is: 
$Q_1 \equiv Q_2 \rightarrow DS(Q_1) \equiv DS(Q_2) $, 
which is
logically equivalent to $DS(Q_1) \centernot \equiv DS(Q_2) \rightarrow Q_1
\centernot \equiv Q_2$ (contrapositive form). Given two queries $Q_1$ and $Q_2$,
w.l.o.g., we can assume $R(Q_1)$ is a multiset (duplicate sensitive) and
$R(Q_2)$ is a set (duplicate insensitive). We consider a tuple $t$ with
multiplicity $i$ in $D$, which is contained in both $R(Q_1)$ and $R(Q_2)$:

\textit{Case 1:} $i=1$. $R(Q_1) \subseteq  R(Q_2) \bigwedge R(Q_2) \subseteq
R(Q_2)$ on $t$.

\textit{Case 2:} $i>=2$. Because $R(Q_1)$ is a multiset and $R(Q_2)$ is a set,
then $R(Q_1) \centernot \subseteq  R(Q_2) \bigwedge R(Q_2) \subseteq R(Q_1)$. This
is because when a tuple appears in the result set, all of its duplicates must also
appear. 



Combining both cases, we get $R(Q_1) \centernot \subseteq R(Q_2) \bigwedge
R(Q_2) \subseteq R(Q_1)$, together with the contrapositive form of
\autoref{def:equiv}, we have $Q_1 \centernot \equiv Q_2$. This completes the
proof. $\square$
\end{proof}

\smallskip
We have proved that duplicate sensitivity is a necessary condition for query
equivalence, then we can follow this theorem to guide our transformations to
preserve the sensitivity for the query mutant. Next, we illustrate how the
transformation is done on a concrete query.

\subsection{Sensitivity-guided Transformation Synthesis}
The goal of our approach is to find more transformations on a query, which
guarantees the equivalence between the given query and the mutant. The key
insight is to first generate many candidate mutants with
~\autoref{thm:dsth}. To achieve it, we first convert a query to
an intermediate form, then apply transformations and last convert it back to a
SQL query. To ease our discussion, we use relational algebra symbols to
represent the operators. Conventionally, we have the following relational
algebra: $\pi$ for projection, $\sigma$ for selection, $\gamma$ for aggregate
and $\delta$ for distinct. Next, we use an example to illustrate our approach.


\begin{figure}[t]
	\centering
	\includegraphics[scale=0.4]{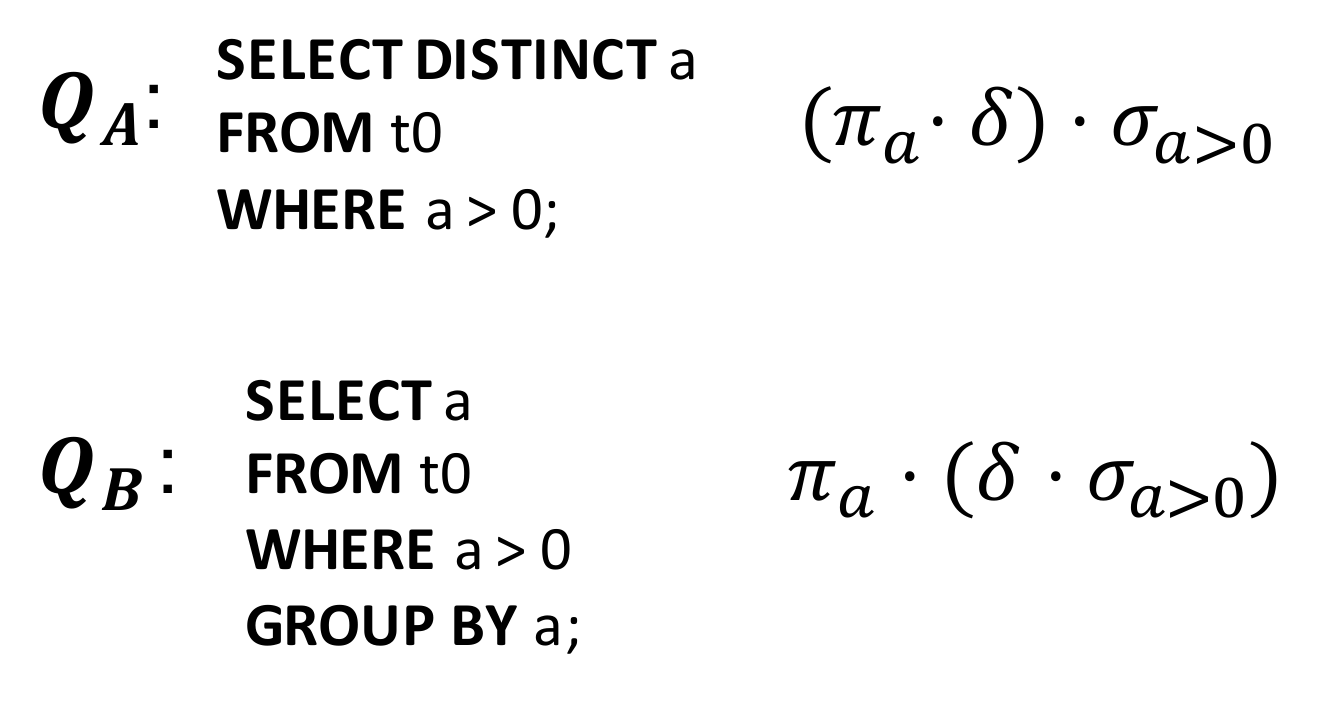}
	\caption{Two operator series of different order have the same execution result. 
	Execution of the relational algebra expressions starts from right to left.}
	\label{fig:trees}
\end{figure}

\begin{table} 
	\centering
	\caption{Remapping from relational algebra to SQL.}
	\label{tb:mapping}
	\begin{tabular}{l l}
		\toprule
		\textbf{Operator} & \textbf{SQL Keywords} \\
		\midrule
		$\sigma$ & HAVING, WHERE \\
		$\delta$ & DISTINCT, GROUP BY \\
		\bottomrule
	\end{tabular}
\end{table}

\subsubsection{A taste}
Let us consider Category 1 in \autoref{fig:motiv} again, viewing the query as a
series of operators. We could obtain an execution order of each query as in
\autoref{fig:trees}.
$Q_A$ performs selection ($\sigma$), de-duplicating ($\delta$) and at last
projection ($\pi$). Among those operations, only the second one changes
multiplicity of every tuple, while the first one keeps or removes all duplicates
and the third one has no impact on the duplicates. However, the problem is how
we can preserve
the duplicate sensitivity while transforming $Q_A$ to $Q_B$. We achieve this by
considering the context of an operator, which means if it is compatible with
both the input and output types of the previous and the next operator.

\subsubsection{In action}
We have three steps to transform a query and to generate a duplicate-sensitivity
preserving mutant. In the first step, we represent $Q_A$ in a relational
algebra expression as $\pi \cdot \delta \cdot \sigma$
(\texttt{SELECT-DISTINCT-WHERE}). In the second step, we apply transformations
on the expression. We have many choices for the transformation, e.g. replace an operator, change
the order of the operators. However, to preserve duplicate sensitivity of the query, we need to
choose the transformation carefully. For instance, if we change an operator from
sensitive to insensitive, it may also alter the sensitivity of the query
(~\autoref{eqtime}). We should always stick to the goal that we should not alter a query's
sensitivity during our transformations. 

Considering the existing transformation 
rules~\cite{20bookconcept}, we apply associativity to the expression and obtain 
two new expressions: $\pi \cdot
(\delta \cdot \sigma)$ or $(\pi \cdot \delta) \cdot \sigma$, which both are still
duplicate sensitive. On the SQL level we can
interpret those expressions differently (\autoref{tb:mapping}. For the first expression, we take out the projection
and we obtain \texttt{SELECT-(DISTINCT-WHERE)} (pull out projection). For the
second expression, we take out the selection and we obtain
\texttt{(SELECT-DISTINCT)-WHERE} (pull out selection). We can also apply
commutativity rules to the expression, then we obtain $\pi \cdot (\sigma \cdot
\delta)$ which we interpret
as \texttt{SELECT-(WHERE-DISTINCT)} on the SQL level. In the last step,
we remap the interpretation to concrete SQL queries. For instance, the
expression \texttt{SELECT-(WHERE-DISTINCT)} is remapped to \texttt{SELECT a FROM
t0 WHERE a>0 GROUP BY a}. We could use \texttt{GROUP BY} instead of
\texttt{DISTINCT} operator because both operator are compatible to the context
following our new definition.
%


\smallskip
To wrap up, our query transformation approach has three steps as the following:

\begin{enumerate}
	\item Convert a SQL query $Q$ to a relational algebra expression $E$;
	\item Apply transformation rules such as associativity and commutativity to
	$E$ and obtain an duplicate-sensitivity preserving expression $E^\prime$.
	The existing rules used for query optimization~\cite{20bookconcept} are also
	applicable.
	\item Remap $E^\prime$ back to a SQL query by considering duplicate
	sensitivity of the SQL operators.
\end{enumerate}


\subsection{Detecting Bugs with Eqsql}
We now could use the duplicate-sensitivity guided transformation to create tests. Following our approach, we designed a tool called Eqsql. We describe
 its architecture for bug finding and the query transformation algorithm in this
 section. We give more implementation details in \autoref{sec:impl}.

\begin{figure}[t]
	\centering
	\includegraphics[scale=0.45]{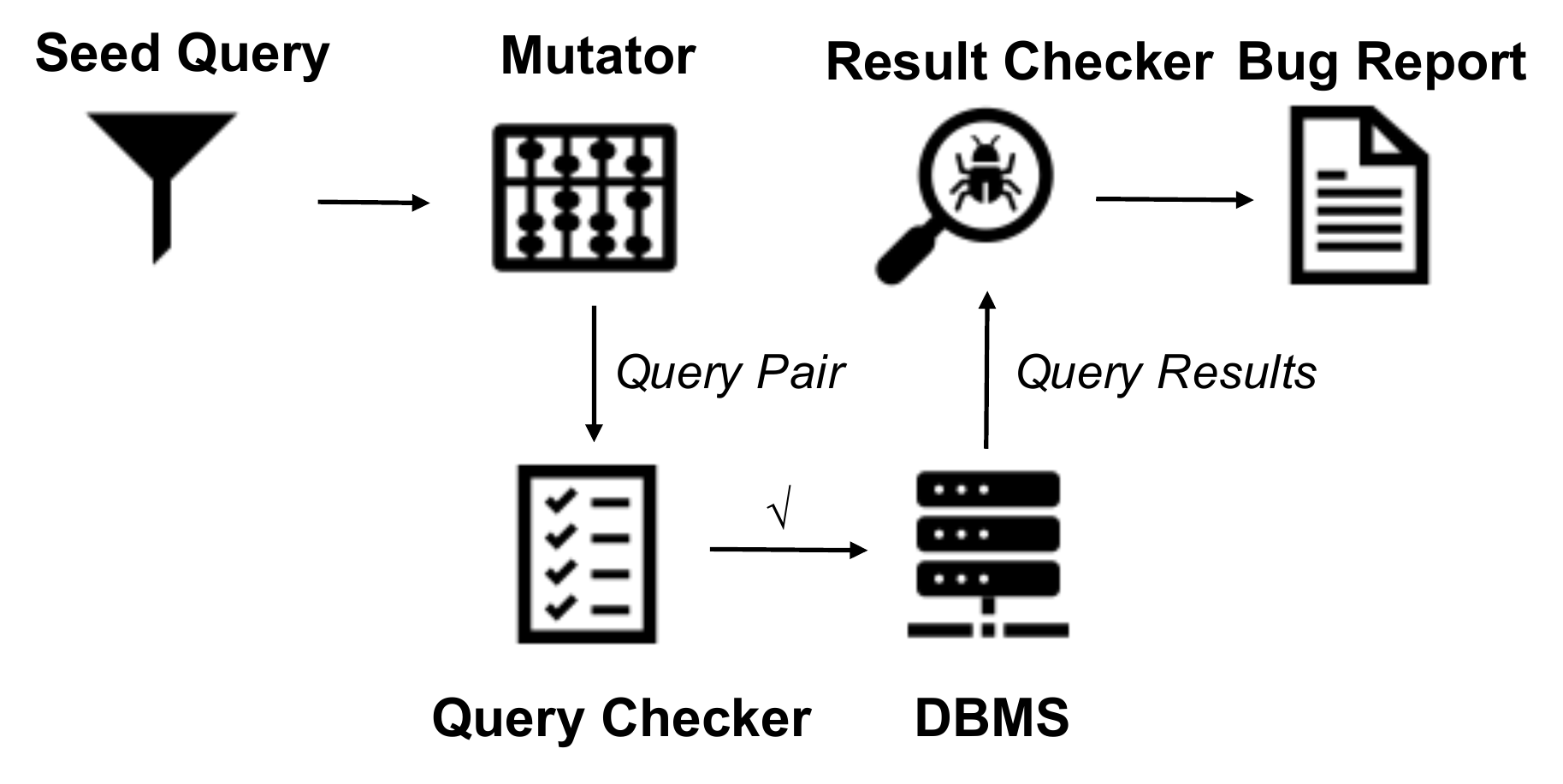}
	\caption{Workflow of guided transformation synthesis}
	\label{fig:workflow}
\end{figure}

\smallskip{\textbf{System Overview}}. \autoref{fig:workflow} shows the
architecture of Eqsql. It takes a bunch of seed queries and processes them one
by one. The Mutator takes a seed query and generates an mutated query pair. The
pair is sent to an equivalence checker where it is discarded if they are nonequivalent.
The equivalent query pairs are executed by a database server and the
results are collected. The result checker reports a bug if it finds a difference in the
pair's result set.

\begin{algorithm}[ht]
	\caption{Eqsql's main logic}
	\label{algo:main}
  \KwIn{A set of randomly given seed queries} \KwOut{A bug report or nothing}
  \SetKwFunction{TestDB}{test\_db} \SetKwProg{Proc}{Procedure}{}{}
  \Proc{\TestDB{$Seeds$}}{\ForEach{$Q \in Seeds$} { \label{algo1:for}
  $Q_e \leftarrow \textsf{transform\_query}(Q)$ \;
  \label{algo1:transform} $checkRes \leftarrow
  \textsf{equal\_checker.check}(Q,Q_e)$ \;
  \label{algo1:checkstart} \If{$checkRes$ is False} {continue \;
  \label{algo1:checkend}} $R \leftarrow \textsf{db.execute}(Q)$ \;
  \label{algo1:result1} 

  $R_e \leftarrow \textsf{db.execute}(Q_e)$ \;
  \label{algo1:result2} 
  
  \If{$R$ is not equal to $R_e$} {\label{algo1:check} \text{report
  a bug}  \; \label{algo1:report}}}}
\end{algorithm}

\smallskip{\textbf{Algorithms}}. \autoref{algo:main} presents the high-level
process of detecting correctness bugs with our approach. It first takes a set of
seed queries $\textit{Seeds}$ and a DMBS under test as input, where the
execution result of a single query is previously unknown during our testing. At
the start of Eqsql's main bug checking logic, we suppose that a bunch of seed
queries are given as input. It iteratively takes a seed query
(\autoref{algo1:for}), and transforms it to get a query mutant
(\autoref{algo:transform}). Next, it sends the query and the mutant to the query
checker where the nonequivalent query pairs are discarded
(\autoref{algo1:checkstart}-\autoref{algo1:checkend}). Then it runs the
equivalent query pairs in the tested DBMS and save the results respectively
(\autoref{algo1:result1}- \autoref{algo1:result2}). At last, it checks if both
queries have the same execution results (\autoref{algo1:check}), if not then it
reports a correctness bug (\autoref{algo1:report}).

\begin{algorithm}[t]
	\caption{Transformation synthesis}
	\label{algo:transform}
	\KwIn{A random seed query $Q$} \KwOut{An EQP $(Q, Q_e)$}
	\SetKwFunction{TestDB}{\textsf{transform}\_\textsf{query}}
	\SetKwProg{Proc}{Function}{}{} \SetKw{KwGoTo}{go to} \SetKw{Break}{break}
   \Proc{\TestDB{$Q$}}{
	  $t \leftarrow \textsf{parse}(Q)$ \; \label{algo2:parse} 
  
	   \ForEach{$r \in rules$}{ \label{algo2:rulestart} \If{$r$ is appliable on
		   $t$} {$t_e \leftarrow \textsf{r}(t)$ \; generate $Q_e$ from $t_e$
		   \; \label{algo2:ruleend} \Break \; }} \Return
		   $(Q,Q_e)$ \; \label{algo2:return}}
  \end{algorithm}

In \autoref{algo:main} we omit that how the seed query is transformed for
simplicity, now we present the procedure \verb+transformQuery+ in
\autoref{algo:transform}. The procedure first 
parses a given query and obtains the corresponding tree representing the algebra
operators (\autoref{algo2:parse}). Then the tree goes through mutants
(\autoref{tb:rules}) to search for an applicable transformation
(\autoref{algo2:rulestart}-\autoref{algo2:ruleend}). At last the procedure
returns the new EQP (\autoref{algo2:return}).

\begin{table*}
	\centering
	\caption{Example relational algebra transformation rules. One row in the
	 table describes a rule. The second column is the relational algebra
	 expression $E$ for a seed query, the third column is a
	 duplicate-sensitivity preserving expression $E_e$ to the second column, the
	 fourth column is a candidate mapping from $E_e$ to a SQL query, and the
	 last column is the name of this rule.}
	\label{tb:rules}
	\begin{tabular}{l  l  l  l  l}
		\toprule 
		\textbf{No.}	&  \textbf{Expression $E$}	& \textbf{Preserving $E_e$}	&
		\textbf{SQL pattern} & \textbf{Name} \\
		\midrule
1 & $\pi \cdot \sigma$  & $\pi \cdot (\sigma)$  & SELECT-(WHERE)  & projection
pull up \\
2 & $\sigma_2 \cdot \sigma_1$  & $\sigma_1 \cdot \sigma_2$  & WHERE-(WHERE)  &
commutative selections \\
3 & $\pi_1 \cdot \pi_2 \cdots \pi_n$ & $\pi_1$ & SELECT & cascade of projection
\\
4 & $E_1 \cup E_2$ & $E_2 \cup E_1$ & SELECT-UNION-SELECT & commutative set
union \\
		\bottomrule
	\end{tabular}
\end{table*}

\section{Implementation}
\label{sec:impl}
We implemented the main
logic of Eqsql including transformations of the
input seed query and result equivalence checks with Go~\cite{wb-golang}. We dispatched the
candidate equivalence checks to a formal query equivalence checker
EQUITAS~\cite{zhou2019automated}. We present the important implementation details
in this section.

\smallskip{\textbf{Seed Generation}}. We leverage the existing tool
go-randgen~\cite{wb-randgen} to randomly generate databases and queries as the
input. go-randgen is a general testing tool following the MySQL SQL
dialects. We manually specify the grammar by providing a Bison YY
file~\cite{wb-bisonyy}.
To use Eqsql to test other DMBSa, only minimum effort is required to set up the
input generator (e.g. using PostgreSQL PHP Generator~\cite{wb-php}).

\smallskip{\textbf{Query Filtering}}. Though we try to precisely specify the correct SQL grammar
for testing, it is stil not easy to get all queries executable because SQL
is a Context-Free Language~\cite{2017hollingum}. With the grammar file, only the
syntax is valid, but the semantics is determined at the runtime. To run the queries,
Eqsql firstly loads all the queries and then attempts to execute every and each of them one by one. To
filter the invalid queries, we created a list
of errors to be ignored. When an error from the list is returned by by the tested DBMS, Eqsql discards
the query immediately. For other queries which may throw an unseen error, it is
possible that a bug is associated with the query. In this case, we keep the query and need manual effort to verify it is a real bug. Due to the given
grammar's complexity, we have an average of 10\% of valid queries observed.

\smallskip{\textbf{Selected Oracles}}. Because of the extensibility of our approach to
generating query candidates, we cannot exhaustively implement all the
possible transformations on the seed query. We selected two transformations for our implementation, while the others would be easily added.  We chose
to implement for \texttt{WHERE} to be consistent with most of the existing works and 
implement for \texttt{GROUP BY} and aggregate functions, which are the magic touch of our
approach.
To obtain the query candidate, a query is first parsed by
sqlparser~\cite{wb-sqlparser} to an Abstract Syntax Tree (AST) which then goes
through transformations. The implementation could be extended to concatenate
different transformations, while our implementation transforms each query only
once.

\smallskip{\textbf{Testing Iteration}}. Each iteration of Eqsql's execution starts by generating the
databases and 2000 queries by default. It first cleans the given database and creates a new one, and then it invokes the main logic to run and verify each of the queries. It saves the query when a bug is discovered. From our experience, such an
iteration takes around 10s to finish when testing against MySQL 8.0.21.


\section{Evaluation} \label{sec:eval}
We guide our quantitative evaluation by the following three research questions:
\begin{itemize}
	\item \textbf{RQ1}: Can Eqsql detect correctness bugs in production level
	databases?
	\item \textbf{RQ2}: What are the impacts of the detected bugs?
	\item \textbf{RQ3}: Can Eqsql improve the code coverage of the input seed queries?
\end{itemize}

\smallskip
\textbf{Results Summary}. We conducted a \textit{two-month} real-world
evaluation of Eqsql's bug finding capability on four MySQL compatible DBMSs: MySQL 5.7, MySQL 8.0, TiDB and CynosDB. The results are highlighted as below:
\begin{itemize}
	\item 
	Eqsql has detected 30 new bugs in the tested DBMSs. They are all unique
	and are confirmed by the developers. This emphasizes the effectiveness
	and practicability of Eqsql.
	\item 
	Most of the detected bugs have been confirmed in several versions of the
	same DBMS, and some of them could be reproduced cross different DBMSs.
	\item Almost all the bugs reported in TiDB (12 out of 13) have been fixed. All the
	bugs in CynosDB have been planned to be fixed, and several of detect bugs in
	MySQL are already patched.
\end{itemize}


\subsection{Evaluation Setup}

\smallskip{\textbf{Environment}}. We performed all experiments on a server
with 80 core Intel(R) Xeon(R) CPU E5-2698 v4 @ 2.20GHz and 256GB physical
memory running Ubuntu-18.04. We ran Eqsql in a single thread process on the server.

\begin{table}[t]
	\centering
	\caption{The tested DBMSs are highly ranked on the DB-Engine and have received thousands of stars on GitHub.}
	\label{tb:db}
	\begin{tabular}{l  l  l  l  }
		\toprule
		\textbf{Name}    &  \textbf{Rank}     & \textbf{Stars} &   \textbf{Size
		(LOC) \tablefootnote{Lines of code.}} \\
		\midrule
		MySQL 5.7              & 2                            & 5.4K & 2,965,321 
		 \\
		MySQL 8.0                 & 2                            & 6.4K & 3,610,003
		\\
		TiDB                         & 60                           & 27.5K & 701,491 
		\\
		CynosDB                      & N/A                          & -
		
		& 3,700,000 \tablefootnote{Provided by the developer. }  \\
		\bottomrule
	\end{tabular}
\end{table}

\smallskip{\textbf{Tested Database Systems}}. Although our approach requires
minimum efforts to be plugged into the existing generators, we still need knowledge
to understand different DBMSs and the corresponding SQL dialects. We
chose MySQL compatible databases as our tested systems, because MySQL is the top
open-source RDBMSs in the DB-Engine Ranking~\cite{wb-rank}. We show information
about the tested systems in \autoref{tb:db}.

\emph{MySQL.} MySQL 5.7 and MySQL 8.0 are considered as different systems in our
evaluation due to that they have distinct differences~\cite{wb-version}. We will
also see that a bug in MySQL 5.7 may not be triggered in MySQL 8.0 and vice
versa. This also gives us the reason to consider them separately. If
there are new releases coming, we shifted to the new release for testing.

\emph{TiDB.} TiDB~\cite{wb-tidb} is a relatively new DBMS developed by PingCAP.
It is a distributed system supporting Hybrid Transactional and Analytical
Processing (HTAP) workloads, and it is compatible with the MySQL protocol. Since
there is a long time duration between its official releases, we continue to test the master
branch of the source repository. We cloned the source code from the official
repository and built the system locally.

\emph{Tencent CynosDB (CDB).} CDB \cite{wb-cdb} is a Tencent Cloud product,
which is a MySQL 100\%-compatible enterprise-level database. It is highly
available and distributed, and it supports more than a million queries per
second. Because CDB is not open-source and it is a cloud product, we set up the
test environment with a Tencent VPS and one CDB instance compatible with MySQL
5.7 and tested the database server on the VPS.

\smallskip{\textbf{Baselines}}. We have to admit that a nuts-and-bolts
comparison among the related tools is hard to achieve. For the early
studies~\cite{slutz1998massive,08adusa,veanes2010qex}, the tools
are not available. More recent tool MutaSQL~\cite{20mutasql} is not
open source. Thus, the only available tool is SQLancer~\cite{Rigger2020PQS, Rigger2020TLP,20norec}. 
We omitted comparison with SQLancer for three reasons. Firstly, both tools support SQL grammar in a 
different way which makes it hard to compare fairly. More importantly, it doesn't
take seed input and ends in log explosion when executing queries. Additionally, we designed
our evaluation to run on new versions of the tested database systems, which is
manual-extensive and objective to run both tools and compare the bugs. Although
we did not design a separate experiment, we did evaluate both tools on
coverage. Our experiment result showed that both tools achieved almost the same
coverage when running for 12 hours by controlling the available parameters. Generally
speaking, our approach is orthogonal to the existing works. Therefore, we will
mainly discuss the newly detected bugs and report coverage improvement of our tool.

\subsection{RQ1: Efficacy in Detection of Correctness Bugs}
All the confirmed bugs found by our tool is shown in \autoref{tb:bugs} with more
details on the number of detected bugs and the bug types. All of bugs included in the table have been uniquely verified for each
database.
In total, we discovered 30 new and unique bugs, among which 20 are value-based bugs and 10
are error-based. ``Value-based bugs" are the ones that the query is executable
but returns wrong result sets. ``Error-based bugs" occur when the query is
executed and it throws an error instead of a result set. The DBMSs under tests have a thorough bug fixing process and it takes
weeks to months to see a bug patched and fixed in the following releases.
For instance, 12/13 bugs of TiDB have been fixed up to the time we completed
writing this paper. For CynosDB, we contacted the developer and were
notified that these bugs will be fixed soon.

In the following part, we enumerate the bug information for each of the tested DBMSs.

\begin{table}
	\caption{Confirmed bugs in the tested DBMSs.}
	\centering
	\label{tb:bugs}
	\begin{tabular}{l c c c c}
		\toprule
		\textbf{Name}  & \textbf{Value}  \tablefootnote{Executable query
		produces incorrect value.} & \textbf{Error}  \tablefootnote{Executable
		query throws error.} & \textbf{Total}  & \textbf{Fixed}\\
		\midrule
		MySQL          & 12   & 2   & 14	&	0	\\
		TiDB           & 6   & 7   & 13	&	12		\\
		CynosDB        & 2    & 1   & 3	&	0		\\
		\midrule
		\textbf{TOTAL} & \textbf{20}  & \textbf{10}		& \textbf{30}	& \textbf{12}
		\\
		\bottomrule
	\end{tabular}
\end{table}

\textbf{MySQL}. We submitted bugs via the MySQL bug
tacker~\cite{wb-mysqltracker}. Most of our new bug report were usually confirmed
within 3 days. Our submitted bugs are all assigned as \textit{severity S3
(Non-critical)}, which is just less severe than S2 (Serious) and S1 (Critical).
Because we expected to find correctness bugs, it is very reasonable they are
assigned as S3. We totally submitted 14 reports for bugs detected by our tool,
and all of them got confirmed by the developers. This means our approach
achieves 100\% true-positive on testing MySQL. We list all detected bugs in
MySQL in~\autoref{fig:bugsample}.

\textbf{TiDB}. Because it is easy for TiDB to have the same bug as MySQL due to
its high-compatibility with MySQL, which makes it more difficult to filter the
duplicated bugs. Though, we have removed duplicates from the final result as
much as possible. For the counted bugs, most of them are tagged as
\textit{Major} or \textit{Moderate} while a few are as \textit{Critical}. 12
of the 13 confirmed bugs have been fixed while one is verified. We list all
detected and valid bugs in TiDB in~\autoref{fig:bugsampletidb}.

\textbf{CynosDB}. CDB does not have a publicly accessible bug tracker, we
contacted the developer and reported the bugs directly to them. For our bug
report, 3 unique bugs are confirmed while the remaining are bugs caused by the
underlying MySQL source code on which the tested version of CDB is implemented.
Because CDB is a commercial product, we can only access and test the version
that are available on the product website, so we stopped testing this DBMS after
we found the reported bugs.

\begin{table}[t]
	\caption{The detected bugs in MySQL are listed here. The column ``Bug ID''
		is the corresponding number for the bug in the MySQL Bug Tracker.
	}
	\label{fig:bugsample}
	\centering
	\begin{tabular} {l l l l l }
		\toprule
		\textbf{No.}	& \textbf{Bug ID} & \textbf{Type}  & \textbf{Status}  &
		\textbf{Affected Versions} \\
		\midrule
		1   & 100209 & Value & Verified & 5.7                     \\
		2   & 100270 & Value & Verified & 5.6.48, 5.7.31          \\
		3   & 100301 & Value & Verified & 5.6.48, 5.7.31          \\
		4   & 100375 & Error & Verified & 5.6.48, 5.7.31          \\
		5   & 100443 & Value & Verified & 5.7, 8.0                \\
		6   & 100453 & Value & Verified & 5.7.31                  \\
		7   & 100489 & Value & Verified & 5.6.48, 5.7.31, 8.0.21  \\
		8   & 100670 & Value & Verified & 5.6.48, 5.7.31, 8.0.21  \\
		9   & 100750 & Error & Verified & 5.7.31, 8.0.21          \\
		10  & 100777 & Value & Verified & 5.7.31, 8.0.21          \\
		11  & 100806 & Value & Verified & 5.7.31, 8.0.21          \\
		12  & 100807 & Value & Verified & 5.7.31, 8.0.21          \\
		13  & 100837 & Value & Verified & 5.6, 5.7, 8.0             \\
		14  & 100985 & Value & Verified & 5.6.48, 5.7.31, 8.0.21  \\
		\bottomrule
	\end{tabular}
\end{table}

\begin{table}[t]
	\caption{The detected bugs in TiDB are listed here. The column ``Bug ID'' is
		the corresponding issue number of the bug in TiDB's repository on GitHub.
	}
	\centering
	\label{fig:bugsampletidb}
	\begin{tabular}{llll}
	\toprule
	\textbf{No.} & \textbf{Bug ID} & \textbf{Type} & \textbf{Status}  \\
	\midrule
	1            & 18314           & Error         & Fixed           \\
	2            & 18493           & Error         & Verified        \\
	3            & 18525           & Value         & Fixed           \\
	4            & 18652           & Error         & Fixed           \\
	5            & 18653           & Error         & Fixed           \\
	6            & 18674           & Value         & Fixed           \\
	7            & 18700           & Value         & Fixed           \\
	8            & 19986           & Value         & Fixed           \\
	9            & 19992           & Value         & Fixed           \\
	10           & 19999           & Value         & Fixed           \\
	11           & 20001           & Value         & Fixed           \\
	12           & 20003           & Value         & Fixed           \\
	13           & 20295           & Error         & Fixed           \\
	\bottomrule
	\end{tabular}
	\end{table}

\subsection{RQ2: Significance of the Bugs}



We studied the impact of the correctness bugs found in different releases of
MySQL. We did not perform this experiment on CDB since only a fixed version
of it is accessible in the cloud. \autoref{fig:bugsample} also shows the number
of reproducible bugs on each major release. We use the milestone releases 5.6,
5.7 and 8.0 to represent all the later releases, and we consider each milestone
as an individual version. The affected versions are verified by the developer(s)
who is assigned to process the bug report.

\smallskip
\textbf{Affected Versions}. Our tool detected 14 bugs in total in MySQL. Among
these bugs, 4 of them could be reproduced on all 3 versions, 7 of
them on 2 versions and 3 of them on 1 version. We evaluated our tool by
testing 5.7.31 and 8.0.21 versions separately. The results show the detected
bugs by our tool have unexpected high impacts on in-production MySQL servers.
Those long-latent bugs have been missed by the regularly executed MySQL Test
Framework~\cite{wb-mysqltest}.It highlights the significance of our bug
finding results. 


\smallskip
\textbf{Feedback From the Developer}. The developers of the tested DBMSs appreciated 
  our bug finding efforts and were amazed by the results.
  Some of them were even curious and requested to use our tool with comments
  like ``how did you find the bugs?'' and ``what method do you use'' because of
  Eqsql's good efficacy in bug detection. After we submitted many bug reports to
  TiDB and shared our experience in DBMS testing, they expressed their
  appreciation for our testing efforts to make TiDB better: \textit{``It is very
  significant that your work has helped us find many bugs in TiDB, while most of
  the bugs are related to correctness. Since correctness is the most important
  thing in the DBMS, it helped us a lot to make TiDB stable and
  resilient.''}

\subsection{RQ3: Coverage Improvement} 
In this section, we show the effectiveness of our approach in improving source code
coverage. We designed our experiment to examine the overall coverage improvement
of Eqsql during a 12-hour testing. This shows how our tool works in practice
with randomly generated test cases. Among the selected DBMSs, we exclude CDB since 
its source code is not available and it is accessible as a cloud
service. For TiDB we have
encountered an issue with the storage which crashes the database server constantly before we
finish our experiment, so we also excluded it from this experiment.


We standardized our experiment with a Dockerfile to automatically set up all
necessary servers and tools. To collect coverage of a running MySQL server, we
compiled and installed MySQL 8.0.21 from source code. The binary is compiled with Gcov
\cite{wb-gcov} support to save runtime and accumulative coverage statistics.
We cleaned all coverage files generated during the installation and the preprocess
phases before we started to run our tool.

We let Eqsql run for 12 hours, then collected the final coverage (Final). Next,
we restarted the server to collect the dry coverage (Dry) with the saved seed
queries. We collected the coverage of criteria line, branch and function.
We report both the coverage by percentage and the absolute increment amount.

\begin{table}[t]
	\caption{Coverage by the seed queries and the queries  after transformations.}
	\label{tb:cov}
	\centering
	\begin{tabular}{c  l  l  l  l}
		\toprule
		\textbf{Name} & \textbf{Type} & \textbf{Line}  & \textbf{Branch}  &
		\textbf{Function}  \\
		\midrule
		\multirow{4}{*}{MySQL} & Dry &  20.8\%  &  11.1\%  &  21.8\%  \\
		                       & Final &  23.4\%  &  12.8\%  &  24.8\%  \\
		                       & \textbf{$\Delta$} \tablefootnote{Increased
		                       coverage computed by $Final-Dry$.}&  +2.6\%  &
		                       +1.7\%  & +3.0\% \\
	                           &\textbf{$\vert \Delta \vert$}
	                           \tablefootnote{Absolute increased amount.} &
	                           +13,848 & +10,075 &  +1,437 \\
		\bottomrule
	\end{tabular}
\end{table}


\autoref{tb:cov} shows the coverage data for MySQL. We can see that after our
transformation, the coverage has increased for each of the criteria
respectively. On MySQL, our tool achieved a 2.6\% coverage improvement for
executed lines, 1.7\% for branches and 3.0\% for functions. Though the increased
coverage counted in percentage is relatively small, our testing has covered more
than ten thousands of new lines and new branches in the source code and also
executed more than one thousand new functions.

\section{Threats to Validity}
Construct validity is concerned with the issue of whether random testing is enough
to demonstrate bug detection capability of our tool. We use a random database
and query generator to test the DBMS, depending on the generator the database
and query may not represent the overall distribution, they should be generated
to maximize features of the DBMS. 

The threats to internal validity lie in the online query equivalent checker.
Because we need to filter the invalid candidates during runtime, it depends on
the checker to provide the result. Our query generation speed is largely limited
by the capability and overhead of the checker. To reduce the threats, we
implemented fixed transformations inside our tool.

External validity deals with issues that limit the ability to generalize our
results. In our experiment, we have used MySQL-compatible DBMS for evaluating
the effectiveness of our approach. However, because our approach does not rely
on specific features of the tested DBMS and it is not designed for a specific
DBMS, our approach should still work on other DBMSs.
\section{Discussions}

{\textbf{Findings}}. Except for what we discussed in ~\autoref{sec:eval}, we also
have interesting findings. Firstly, we found one of the detected
bug~\cite{wb-mulbug} in MySQL can be reproduced in MariaDB and TiDB. It is
reasonable that such a bug exists in MariaDB~\cite{wb-mariadb}, which is a fork
of MySQL. However, it is unexpected that the bug exists in TiDB which is an
independently-developed DBMS. We had a brief discussion with a TiDB developer on
this bug. He mentioned that the bug may be introduced when implementing the
MySQL protocol with referencing to the original source code. This implies that
more such bugs can probably exist in TiDB's underlying source code and in other
DBMSs which reference MySQL source code in a similar way. Secondly, our tool
detected a parsing bug~\cite{wb-docbug} in MySQL, which was firstly confirmed by
the MySQL verification team but was classified as not a bug after a few hours.
The reason for this update was because the team has found the bug is claimed as
a documented defect, though the test case has no violation with the SQL
standard. This is an example of how a DBMS's implementation can deviate from the
standard.


\smallskip{\textbf{Extensions}}. Currently, we only implemented two fixed
transformations for MySQL-compatible systems. To run our tool for other DBMSs,
two main components are required, i.e., a connector of the specific server and a
SQL grammar file. A Golang native connector is easier to use with our tool but
many new systems do not provide such support. In this case, our tool can be used
as a seed generator and the output queries could be exported to fit into the
testing process. The grammar file is customizable on any level to the user's
need, however manual efforts are required to understand and write the file.

\section{Related Work}\label{sec:related} 
In this section, we discuss the
approach for testing DBMS with an emphasis on correctness bugs. Particularly, we
show how the related works solve the problem of deciding the expected result set
of a query.

\smallskip{\textbf{Constraint Solving}}. A previous common solution to generate
the expected result of a given query is to use constraint solving. ADUSA~\cite{08adusa,abdul2010automated} generates test oracle by translating a
SQL query to an Alloy \cite{wb-alloy} specification, which describes the
constraints in a query. By solving the constraints with an SMT solver, it knows
which concrete values can satisfy the constraint and should be returned by the
original query. It then inserts the value to the DBMS as data and checks if
executing the query will return the same value. However, due to the incomplete
model, ADUSA can only convert query with limited data types to Alloy
specification. For example, it cannot convert a DBMS varchar field into the
Alloy language. Qex ~\cite{veanes2009symbolic,veanes2010qex} is another
symbolic query explorer which models more SQL features, but its practical
adoption is hindered by high overhead in symbolically processing big tables and
complex joins.

More recently, PQS~\cite{Rigger2020PQS} synthesizes queries instead of solving
constraints.
 It implements an expression generator, which takes a table row
from the database and generates an expression that is evaluated to true on the
row. Then it runs the query in the DBMS and checks if the row is contained in a
query using the generated expression as the ``where'' condition. Though PQS can
generate the test oracle more practically than the previous work, it can only reveal
bugs with a symptom of missing the picked row. Compared to all the previous
works, our idea is more generally applicable to the DBMS, and it can detect bugs
having different symptoms (e.g., ``order by'' has wrong ordering).

\smallskip{\textbf{Differential Testing}}.  
Differential testing is a common approach to decide which result set is
incorrect. RAGS~\cite{slutz1998massive} and SparkFuzz~\cite{ghit2020sparkfuzz}
are in this category. They compare the result sets of one query from several
chosen DBMSs, which are expected to return the same result set. The SQL Server
team proposed a self-differential approach~\cite{09microsoft} for their query
optimization rules, which compares results of the same query with turning on and
off the optimization. However, differential testing cannot be used to test
unique extensions such as PostgreSQL's data type \texttt{path} to save a
geometric path on a plane~\cite{wb-pgtype}. 

\smallskip{\textbf{Metamorphic Testing}}. The key idea of metamorphic testing is
to detect output violations of a bunch of related test
cases~\cite{02metamorphic}. In the context of our problem, it means to derive
from the existing queries to obtain query mutants that have known result sets.
Similarly, NoREC~\cite{20norec} constructs a non-optimizing version of a query
and compares the result. However, they are only limited to testing the
optimization, while our approach generally tests the DBMS.
MutaSQL~\cite{20mutasql} takes a given equivalent transformation and construct a
pair of equivalent queries and reports a bug when they return different result
sets. In comparison, our approach use guided synthesis to discover new equivalent
transformations, not relying on developer knowledge to provide them.

\section{Conclusion}
In this paper, we presented a fully automated approach to generate equivalent query pairs, namely Eqsql,
which can detect correctness bugs in the DBMS. Eqsql's main contribution is a necessary condition for query
equivalence, which is leveraged to generate many candidate query mutant. To evaluate the efficacy of Eqsql,
we have used it to test several production-level DBMSs including MySQL, TiDB and CynosDB. Our evaluation results
show that Eqsql can detect in total 30 bugs which are all uniquely confirmed . 

\balance
\bibliographystyle{IEEEtran}
\bibliography{mybib}

\end{document}